
\input phyzzx
\hsize=6in
\vsize=8.5in
\nopagenumbers

\title{Spin Connections and Classification of Inequivalent Quantizations}

\author{Kazuhiko ODAKA}
\centerline{Department of Mathematics and Physics}
\centerline{National Defence Academy}
\centerline{Yokosuka, 239 JAPAN}

\vskip 5mm
We discuss an extention of the quantization method based on the induced
representation of the canonical group.

\vskip 10mm
Quantum mechanics on a general configuration space was firstly studied
by Dirac$^1$. His method is the base of the quantum investigation of
constraint systems. Next, Mackey$^2$ proposed another quantization method.
He take a homogeneous space as the configuration space and he used the induced
representation theory of group developed by Wigner$^3$. The interesting point
of his study is to show that there exist many inequivalent quantizations for
general configuration space cases and that Dirac's method is one of them.
But, Mackey's method is not so general, because the configuration space is
the homogeneous space. Then, we will study the inequivalent quantization
problem from more general viewpoint.

We consider a wavefunction as the most fundamental object in the quantum
theory. Thus, we define firstly the wavefunction ($\psi(q)$) over the
configuration space(Q) consistently and specify its properties including
time-evolution. The wavefunction is not observed directly and a physical
observable is a probability density ($\rho(q)$), which is defined as
$\rho(q)\equiv\psi(q)^\dagger \ast\psi(q)$. The wavefunction is assumed to be
a n-component complex valued function. Then, $\dagger$ means complex conjugate
and $\ast$ is inner product.

Next we consider the time(t) evolution of the wavefunction.  The probability
density should satisfy the equation of continuity;
${d\over dt}\rho(q t)~=~-{\partial \over \partial
\theta^a}J^a(q t)$, where $\theta^a$ is locally orthogonal coordinates.
A new physical quantity $J^a(qt)$(probability current density) must be
introduced for the probabilistic interpretation of the wavefunction. The
form of the current must be determined and so we introduce the equation
of the time evolution of the wavefuction, which is a linear first differential
equation for time; ${d \over dt}\psi(qt)~=~\hat H(q) \psi(qt)$ because of the
probabilistic interpretation and of the principle of superposition.
The forms of $J^a$and $\hat H$
are restricted to
$$\hat H(q)~=~i{\partial \over \partial \theta^a}C(q)
{\partial \over \partial \theta^a}~+~V(q),~~J^a(qt)~=~C(q){\partial \over
\partial \theta^a}\psi(qt)^\dagger *\psi(qt)~+~C.C.$$
where $C(q)$ is a undetermined function. There exists another posibility
(Dirac equation like) but the following arguments are not changed.

The inner product is written as $\rho(qt)\equiv\psi(qt)^{\mu\dagger}G_{\mu\nu}
(q)\psi^\nu(qt)$, where $G_{\mu \nu}(q)$ should satisfy the properties of the
metric. Thus, we can study the transformation properties of the wavefunction
in imitation of Riemannian geometry. The correspondnce relations are $\vec
e_\mu \equiv {\partial \over \partial x^\mu}~\iff~\vec E_\mu(q),
{}~\vec E_\mu^\dagger(q),~~~
\vec V~=~V^\mu \vec e_\mu~\iff~\psi(qt)~=~
 \vec E_\mu (q) \psi^\mu(qt),~~~(\vec e_\mu,\vec e_\nu)=g_{\mu \nu}~\iff~
(\vec E_\mu^\dagger,\vec E_\nu) = G_{\mu \nu}(q)~and~(\vec V,\vec U) = V^\mu
U^\nu g_{\mu\nu}~\iff~(\psi^\dagger,
\psi) = \psi^{\dagger\mu}\psi^\nu G_{\mu\nu}$ .

Now, we assume that in the locally orthogonal coordinate systems the metric of
the wavefunction is Kronecker's delta. But there are many choices of the
locally orthogonal coordinate systems and  these systems are related each
other by rotation. The wavefunction transforms with these rotations but
physical observable ($\rho$) should be independent of the choices of the
systems. Therefore, the wavefunction must be the base of the representation
of the rotation group. This means the introduction of the spin degree.

Let us study the transformation property of the other physical observables.
The definitions of the current and Hamiltonian include the derivative. Thus,
we need to introduce the connection coefficients for the locally rotation.
The basis vectors of the locally orthogonal coordinate system are dependent
on a position. They change under the transfor of position as
$\nabla_{\vec e_a} \vec e_b(q)=\Gamma_{ab}^c \vec e_c(q)$  where
$\Gamma_{ab}^c$ is affine connection coefficients. Therefore, the basis
vectors at a near point is given by  $\vec e_b(q+d\theta^a\vec e_a)~=~
(\delta_{bc}+d\theta^a\Gamma_{ab}^c)\vec e_c(q)$. This part
$(\delta_{bc}+d\theta^a\Gamma_{ab}^c)$ means a infinitesimal rotation.
Corresponding to this infinitesimal rotation, the base of the wavefunction
transfors as $\vec E_\mu(q+d\theta^a\vec e_a)~=~V_\mu^\nu \vec E_\nu(q)$
where $V_\mu^\nu$ is the unitary representaion of the infinitesimal
rotation. Then, the spin connection is given by $$\tilde \nabla_{\vec e_a}
\vec E_\mu~=~Tr[\Gamma_{ab}^c (t^\alpha)_c^b] T^{\alpha \nu}_\mu \vec E_\nu $$
where $t^\alpha$ is an ajoint representaion of the so(d) Lie generator and
$T^\alpha$ is the n dimentional representaion.
We introduce an one-form $d\theta^a A_a^\alpha~\equiv~d\theta^a
Tr[\Gamma_{ab}^c (t^\alpha)_c^b]$ which transforms as $A^\alpha
T^\alpha~\Longrightarrow~-~
V^\dagger dV + V^\dagger A^\alpha T^\alpha V$ under the local rotation
$v$. Here $V$ is the unitary representation matrix of $v$. The current
$d\theta^aJ_a~=~-\psi^{\dagger \mu}[\delta_{\mu \nu}d-A^\alpha
T^\alpha_{\mu\nu}]\psi^\nu + C.C.$
is invariant under this rotation.

Next problem is to define the locally orthogonal coordinate systems over the
manifold that is the configuration space $Q$. This problem is a pure
mathematical problem. We introduce some charts and the locally coordinate
are defined on each chart. In the overlap region two coordinate systems are
consistently connected. That is, we define single group-valued function over
the overlap region. When the overlap region is $S^d$,  the single
group-valued function are classifyed by the homotopy group. For examples,
$\Pi_3(SO(d))=Z$ and $\Pi_1(SO(2))=Z$.
Thus, the connection coefficients can be classifyed and it may be expected
that quantization methods are done according to this claasification. But
some cases of these classes may be unitary equivarent. We have not
established this situation.

Let us apply our idea to the d-dimentional sphere cases. We immerse a sphere
in $R^{d+1}$ space, introduce two charts and use the stereographic projection.
In each charts, basis vectors are given by $\vec e_a~=~
{1 \over 1+ \vec x^2}\big(\delta _{ai}(1+\vec x^2)-2x_a x_i,
{}~2x_a\big)$ and $\vec e_a~=~
{1 \over 1+ \vec z^2}\big(\delta _{ai}(1+\vec z^2)-2z_a z_i,
{}~-2z_a\big)$, and the connection coefficients are

$$(S)~~~~{1+\vec x^2 \over 2}{\partial \over \partial x_a} \vec e_b~=~
\Gamma_{ab}^c \vec e_c,~~~\Gamma_{ab}^c~=~(\delta_{ab}x_c-\delta_{ac} x_b)$$

$$(N)~~~~{1+\vec z^2 \over 2}{\partial \over \partial z_a} \vec e_b~=~
\Gamma_{ab}^c \vec e_c,~~~\Gamma_{ab}^c~=~(\delta_{ab}z_c-\delta_{ac} z_b).$$
We can get the derivative of the wavefunction $$(S)~~
[{1+\vec x^2 \over 2}{\partial \over \partial x_a}~-~Tr[\Gamma_{ab}^c
(t^\alpha)_c^b] T^{\alpha}]\psi,~~~
(N)~~
[{1+\vec z^2 \over 2}{\partial \over \partial z_a}~-~Tr[\Gamma_{ab}^c
(t^\alpha)_c^b] T^{\alpha}]\psi.$$

Finally, for the $S^d(\simeq SO(d+1)/SO(d))$ case we show the relation between
the induced representation approach$^{4,5}$ and ours. The derivative on the
tangent space is constructed from the generators of $SO(d+1)$. This
correspondence is
$$\sum_b {r_b \over r^2} G_{b a}~\iff~{1+\vec x^2 \over 2}{\partial
\over \partial x_a}~-~Tr[\Gamma_{ab}^c (t^\alpha)_c^b] T^{\alpha}$$
where $G_{b a}$ is the generatotrs of SO(d+1) and $\vec r$ is the
position vector on $S^d$ in the $R^{d+1}$ space.

\vskip 5mm
\noindent
References

\noindent
1. P.A.M. Dirac, Lectures on Quantum Mechanics (Yeshiva, New York, 1964)

\noindent
2. G.W. Mackey, Induced Representations of Groups and Quantum Mechanics

\noindent
\hskip 5.2mm
(Benjamin, New York, 1969).

\noindent
3. E. Wigner, Ann. Math. 40(1939)149.

\noindent
4. C.J. Isham, in Relativity, Group and Topology II (ed. B.S. de Witt and

\noindent
\hskip 5.2mm
R. Strora, North-Holland, Amstrerdam 1984),

\noindent
\hskip 5.2mm
N.P. Landsman and N. Linden, Nucl. Phys. B365(1991)121,

\noindent
\hskip 5.2mm
Y. Ohnuki and S. Kitakado, J. Math. Phys. 34(1993)2827.

\noindent
5. K. Odaka, to apear in Proceeding of Symmetries in Science VIII
(Bregenz, 1994).
\bye